# Comment on 'Semiclassical Klein-Kramers and Smoluchowski equations for the Brownian motion of a particle in an external potential'

Roumen Tsekov
Department of Physical Chemistry, University of Sofia, 1164 Sofia, Bulgaria

The range of validity of the semiclassical Smoluchowski equation derived recently by Coffey et al is discussed. The analysis is based on the quantum Smoluchowski equation derived by the present author before. A quantum generalization of the Einstein law of Brownian motion is also obtained.

Recently, Coffey et al [1] have proposed a heuristic method for determination of the effective diffusion coefficient of a quantum Brownian particle. Starting from the master equation for the Wigner distribution function they have derived a semiclassical Smoluchowski equation

$$\partial_t P = \partial_x [P \partial_x V / \zeta + \partial_x (D_{eff} P)] \tag{1}$$

Here $P$ is the probability density of the Brownian particle, $\zeta$ is the friction coefficient, $V$ is an arbitrary external potential and $D_{eff} = D(1 + \beta^2 \hbar^2 \partial_x^2 V / 12m)$ is the effective position-dependent diffusion coefficient, where $D = k_B T / \zeta$ is the classical Einstein diffusion constant and $\beta = 1/k_B T$. It is shown that Eq. (1) provides the correct equilibrium distribution in the frames of the adopted semiclassical approximation. However, Eq. (1) does not describe well enough the evolution. For instance, in the case of a free quantum Brownian particle ($V = 0$) it reduces to the classical diffusion equation without any quantum effect left. Hence, according to Eq. (1) the position dispersion of a free quantum Brownian particle obeys the classical Einstein law $\sigma^2 = 2Dt$. Numerical simulations [2] have shown, however, that the width of the wave packet of electrons exhibits anomalous diffusion with $\sigma \sim t^\alpha$, where $\alpha$ depends strongly on the hopping strength.

A decade ago a Letter to the Editor of the present journal was published [3], where the following quantum Smoluchowski equation is derived

$$\partial_t P = D \partial_x [P \partial_x \int_0^\beta P^{-1/2} (\hat{H} + 2\partial_\beta) P^{1/2} d\beta] \tag{2}$$

Here $\hat{H} \equiv -(\hbar^2/2m)\partial_x^2 + V$ is the Hamiltonian of the quantum Brownian particle. It is proven that the equilibrium solution of Eq. (2) is the quantum canonical Gibbs distribution. In contrast to Eq.

(1), being semiclassical expansion on small $\hbar$, Eq. (2) is general. The corresponding quantum Fokker-Planck and Klein-Kramers equations in the momentum and coordinate-momentum spaces, respectively, are also proposed [4]. In the case of a harmonic oscillator the solution of Eq. (2) is a Gaussian distribution with dispersion satisfying the following equation [3]

$$\partial_t \sigma^2 = 2D(1+\sigma^2 \int_0^\beta \frac{\hbar^2}{4m\sigma^4} d\beta - \beta m\omega_0^2 \sigma^2) \tag{3}$$

The equilibrium solution of Eq. (3) is the well-known expression $\sigma_e^2 = (\hbar/2m\omega_0)\coth(\beta\hbar\omega_0/2)$ from the quantum statistical physics. If someone is interested in the semiclassical limit, one must replace the dispersion in the quantum term above by the corresponding classical expression. If in addition one replaces $\sigma^2$ there by its classical value at equilibrium $k_B T/m\omega_0^2$, Eq. (3) reduces to

$$\partial_t \sigma^2 = 2D(1+\beta^2\hbar^2\omega_0^2/12 - \beta m\omega_0^2 \sigma^2) \tag{4}$$

which coincides with the result of Coffey et al. Therefore, their approximation corresponds to relaxed quantum fluctuations.

The difference becomes more significant in the case of a free Brownian particle, where Eq. (3) reduces to

$$\partial_t \sigma^2 = 2D(1+\sigma^2 \int_0^\beta \frac{\hbar^2}{4m\sigma^4} d\beta) \tag{5}$$

The approach of Coffey et al corresponds to Eq. (5), where in the last quantum term the dispersion is replaced again by its equilibrium classical value being infinity. This is, however, a very rough approximation resulting in complete loss of the quantum term. That is why Eq. (1) reduces to the classical diffusion equation in the case of a free quantum Brownian particle. The term in the brackets represents in fact the relative increase of the diffusion coefficient due to quantum effects. Since the dispersion $\sigma^2$ increases in time, at large times it is large enough that the quantum term, being inverse proportional to $\sigma^2$, becomes negligible. Hence, at large $t$ the solution of Eq. (5) tends asymptotically to the classical Einstein relation $\sigma^2 = 2Dt$. On the contrary, at small $t$ the dispersion $\sigma^2$ is small enough that the quantum term in Eq. (5) is dominant. Hence, one can neglect now the unity in the brackets of Eq. (5) and the solution of the remaining equation is

$$\sigma^4 = \frac{\hbar^2}{m\partial_\beta(\beta\zeta)}t \qquad (6)$$

This is an interesting result showing that the quadrate of the dispersion is linearly proportional to time. It correlates well to the numerical simulations, where $\alpha$ tends to about 0.25 at zero hopping strength (see Fig. 7 in Ref. [2]). The latter limit corresponds to zero temperature in our case, where Eq. (6) reduces to the pure quantum expression $\sigma = (\hbar^2 t/m\zeta)^{1/4}$. Usually the diffusion coefficient in liquids obeys the Arrhenius law, $D = D_0 \exp(-\beta E_a)$, where $E_a$ is the diffusion activation energy. Introducing this model in Eq. (6) yields $\sigma^2 = 2\lambda_E \sqrt{Dt}$ showing that, the quantum diffusion is important at small displacements up to $\lambda_E = \hbar/2\sqrt{mE_a}$. According to this relation, the quantum transport has a typical length scale of the order of the de Broglie wavelength of the diffusion activation energy. This is a new result since it is usually accepted that the length scale of the quantum Brownian motion is the thermal de Broglie wavelength. It is not surprising, however, since a main feature of quantum particles is the tunneling effect through potential barriers.

If one is interested as Coffey et al in the exact semiclassical limit, one must replace the dispersion on the right hand site of Eq. (5) with its classical but non-equilibrium value $\sigma^2 = 2Dt$. After integration on time, one yields

$$\sigma^2 = 2Dt + (D^2 \int_0^\beta \frac{\hbar^2}{4mD^2} d\beta)\ln t + const \qquad (7)$$

As seen there is a logarithmic term proportional to $\hbar^2$, which is missing in the theory of Coffey et al due to suppressed quantum relaxations. In fact, the application of their approach to a free Brownian particle is good only at very large time, which coincides with the classical limit.

One is probably curious what the exact solution of Eq. (5) is. Unfortunately, it is impossible to solve it in general, but if one is interested in strong quantum effects at low temperature, one can neglect the quantum entropic effect. Thus, performing the integration on $\beta$ in Eq. (5) at constant $\sigma^4$ leads to

$$\partial_t \sigma^2 = 2D(1 + \lambda_T^2/\sigma^2) \qquad (8)$$

where $\lambda_T = \hbar/2\sqrt{mk_B T}$ is proportional now to the thermal de Broglie wave length. Note that $\lambda_T$ represent the position uncertainty calculated from the minimal Heisenberg principle for the classical momentum dispersion. The solution of Eq. (8) reads

$$\sigma^2 = \lambda_T^2\{-1-W_{-1}[-\exp(-1-2Dt/\lambda_T^2)]\} \qquad (9)$$

which is expressed via a Lambert $W$-function. In the case of large time Eq. (9) tends asymptotically to the Einstein law, while at short time the purely quantum expression $\sigma^2 = \hbar\sqrt{t/m\zeta}$ holds. Equation (9) can be approximated well by elementary mathematical functions via the expression $\sigma^2 = 2Dt + 2\lambda_T^2 \ln(1+\sqrt{Dt}/\lambda_T)$, which reduces to Eq. (7) in the semiclassical limit. This result differs substantially from the expression derived by Ford and O'Connell [5] but the reason is obvious. The theory of Ford and O'Connell describes the motion of a classical Brownian particle in a quantum environment. This is evident from the fact that in the case of removal of the thermal bath, the remaining equation of motion of their particle is the Newtonian one. The subject of Eq. (2) is just the opposite: a quantum particle moving in a classical environment. Hence, without the thermal bath our particle is described by the Schrodinger equation [3].

Finally, if the system is close to equilibrium one can replace the probability density in the so-called quantum potential in Eq. (2) by the classical Boltzmann distribution $P_e \propto \exp(-\beta V)$ to obtain the semiclassical description of the last stage of the probability evolution. By doing so one can derive from Eq. (2) the following equation

$$\partial_t P = \partial_x [P\partial_x V_{eff}/\zeta + \partial_x(D_{eff}P)] \qquad (10)$$

where $V_{eff} = V + \beta\hbar^2\partial_x^2 V/24m$ is an effective potential. This equation has been already derived by Ankerhold et al [6] and Coffey et al [1] noticed that it is not identical to Eq. (1). The present analysis supports Eq. (10) as the correct semiclassical Smoluchowski equation. The latter, however, neglects also the quantum relaxations, since in the derivation above the equilibrium Boltzmann distribution was employed instead of the non-equilibrium classical one. One could read also the replay of Coffey et al [7].